\documentclass[conference]{IEEEtran}
\IEEEoverridecommandlockouts
\usepackage{cite}
\usepackage{amsmath,amssymb,amsfonts}
\usepackage{nameref}
\usepackage{algorithmic}
\usepackage{graphicx}
\usepackage{textcomp}
\usepackage{xcolor}
\usepackage{prs}

\pdfoutput=1

\def\BibTeX{{\rm B\kern-.05em{\sc i\kern-.025em b}\kern-.08em
    T\kern-.1667em\lower.7ex\hbox{E}\kern-.125emX}}
\begin{document}

\title{Opportunistic Mutual Exclusion

}

\author{\IEEEauthorblockN{Karthi Srinivasan}
\IEEEauthorblockA{\textit{Department of Electrical Engineering} \\
\textit{Yale University}\\
New Haven, USA \\
karthi.srinivasan\symbol{64}yale.edu}
\and
\IEEEauthorblockN{Yoram Moses}
\IEEEauthorblockA{\textit{Department of Electrical Engineering} \\
\textit{Technion-Israel Institute of Technology}\\
Haifa, Israel \\
moses\symbol{64}ee.technion.ac.il}
\and
\IEEEauthorblockN{Rajit Manohar}
\IEEEauthorblockA{\textit{Department of Electrical Engineering} \\
\textit{Yale University}\\
New Haven, USA \\
rajit.manohar\symbol{64}yale.edu}
}
\maketitle
















\begin{abstract}
Mutual exclusion is an important problem in the context of shared resource usage, where only one process can be using the shared resource at any given time.
A mutual exclusion protocol that does not use information on the duration for which each process uses the resource can lead to sub-optimal utilization times.
We consider a simple two-process mutual exclusion problem with a central server that
provides access to the shared resource.
We show that even in the absence of a clock, under certain conditions, the server can opportunistically grant early access to a client based on timing information. We 
call our new protocol opportunistic mutual exclusion.
Our approach requires an extra request signal on each channel between client and server to convey extra information, and the server can grant early access 
based only on the order of events rather than through measuring time.
We derive the handshaking specification and production rules for our protocol,
and report on the energy and delay of the circuits in a 65nm process.

\end{abstract}


\section{Introduction}

The mutual exclusion problem - guaranteeing mutually exclusive access to a certain shared process among a number of other competing processes that each request for access  - has been known for decades, and several algorithms have been proposed to solve this under various models \cite{lamport1987fast,raynal1986algorithms,dijkstraME}.

Conventionally, the mutual exclusion problem is solved by instantiating a central server that holds a token which grants access to the shared resource. When any of the clients request for use, the server hands out the token to the client, which then uses the resource and returns the token to the server once it is done. This process then repeats. Since only one client may hold the token at any time, mutual exclusion is guaranteed. There also exist distributed solutions to this problem---for example using rings \cite{martin1985distributed,martin1990asynchronous}, or trees of processes \cite{raymond1989tree}. 

Generally, no assumptions on the behavior of the requesting processes is assumed in the design of the mutual exclusion server that handles the whole system. However, in the context of asynchronous circuits, timing analysis and timing simulations are used to determine performance during the design flow \cite{hua2017exact,hua2020cyclone}. These measurements can also inform efforts to further optimize the design. In this article, we look at an opportunity to optimize the mutual exclusion process, based on knowledge of the timing behavior of the requesting processes, and present a novel technique to achieve the same. 

We look at the simple case with a single token server and two clients making requests to use a shared resource. We show that if certain bounds on the timing behavior of the clients are known by the server beforehand, then it can, without an internal clock and using only the ordering of signal arrivals from the clients, pre-emptively grant access to one client while another is still using, and thus reduce the idle time of the shared resource, potentially by an unbounded amount. 

In the following sections, we detail the kinds of timing constraints that we use in the design of the opportunistic mutual exclusion circuit, and show how these particular constraints can be incorporated into the circuit itself. We then derive a straightforward implementation of the circuit that uses three arbiters. Next, we show that there is an alternate implementation which seemingly uses two arbiters but can in fact be reduced to a single arbiter. We conclude with SPICE simulation results and a discussion of potential uses of this circuit.  

\section{Timing Forks and Zigzags}

In circuit design, there are ways to infer ordering of pairs of events using a common event that caused them both. This is particularly interesting in situations where it is possible to infer the time ordering of two events on two different processes despite there existing no actual communication between the two processes.These are known as point of divergence constraints. A common example of this is the setup time constraint in synchronous logic design, where the data must be valid at least a certain time before the clock edge arrives. 

\begin{figure}[t]
\centerline{\includegraphics[scale=1.083]{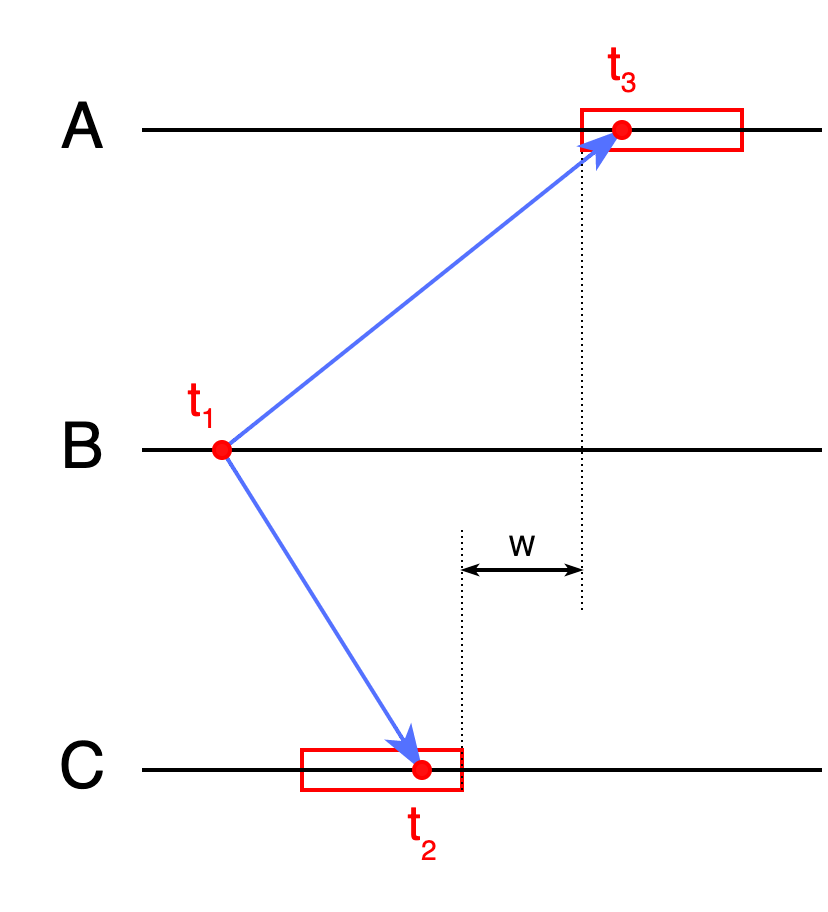}}
\caption{Timing Fork. A simple example of a point of divergence constraint. Red boxes represent the windows of time during which an event may occur. Red dots represent a particular realization of the event. }
\label{fork}
\end{figure}

A simple model of such a point of divergence constraint, known as a timing fork, is shown in Fig. \ref{fork}. Suppose A, B and C are three processes such that an event in B, $e_1$ causes, directly or indirectly, events $e_3$ (called the head) and $e_2$ (called the tail) on A and C respectively. Let the time of occurrence of event $e_i$ be $t_i$. Assume that the delay between $t_3$ and $t_1$ is $d_A$, and that between $t_2$ and $t_1$ is $d_C$, both of which can fall anywhere within a certain range. Now, if $d_A$ is always larger than $d_C$, we can conclude that:
\begin{flalign*}
W &= \inf(d_A - d_C) \\ &= \inf((t_3 - t_1) - (t_2 - t_1)) \\ &= \inf(t_3) - \sup(t_2) > 0
\end{flalign*}

This kind of constraint regarding the relative time difference of two event occurrence times is called a timing fork, and the quantity $W$, which measures the minimum time separation between the two events, is referred to as the weight of the fork. Note that the `minimum separation' intuition only holds for positive-weight forks. If the weight of the fork is negative, then the definition does not change, but the intuition about the quantity it captures switches to being a `maximum separation'. In addition, the three processes do not have to be distinct. There can be a degenerate case where A and B, or C and B, are the same process. The constraint still holds as long the occurrence times still behave the same way. Note that the actual times can occur over a range, but the weight of the fork is calculated based on the upper and lower bounds of these occurrence times.
Now, we can extend this concept by using multiple timing forks as follows.  

\begin{figure}[t]
\centerline{\includegraphics{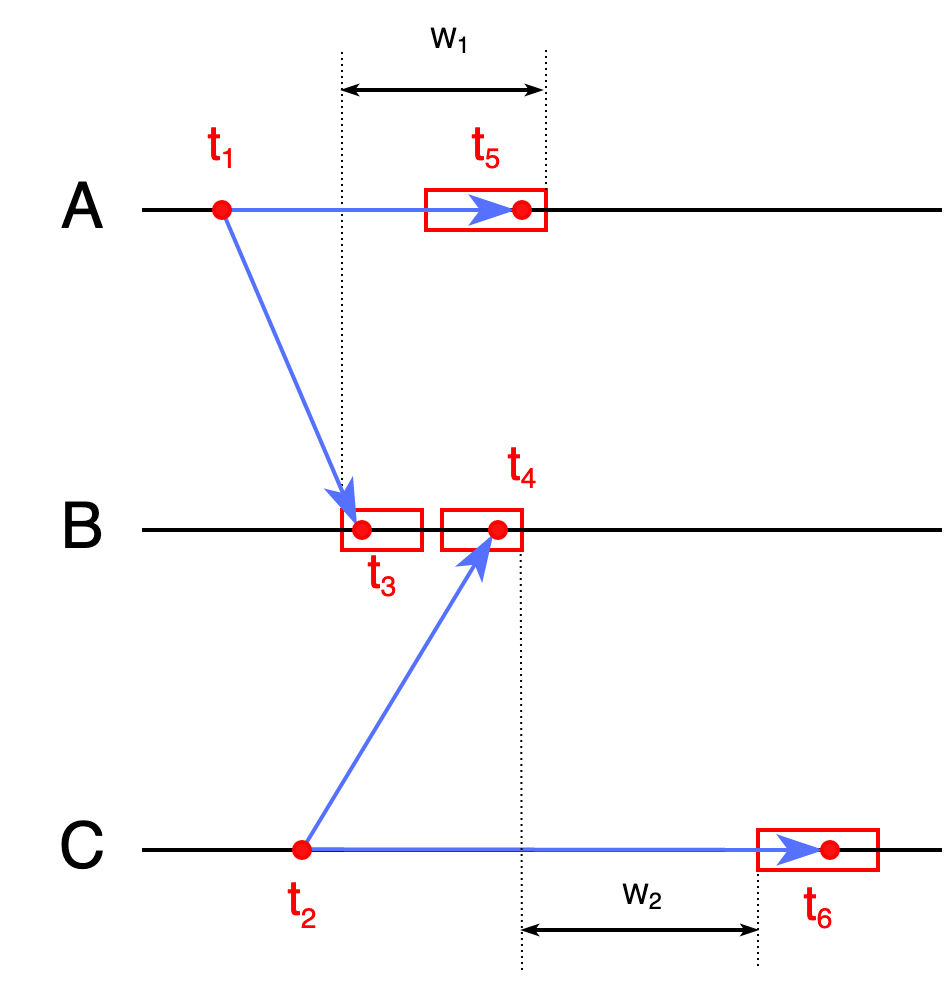}}
\caption{Timing Zigzag. Combining two timing forks, with ordering information on a pair of events to infer the relative time of events that do not share the same point of divergence. }
\label{tzz_exp}
\end{figure}

 Consider the case shown in Fig. \ref{tzz_exp}, where A, B and C are three different processes. We want to make an absolute decision on the ordering of events $e_5$ and $e_6$, without information about the actual ordering of the events that caused them, $e_1$ and $e_2$. 

Suppose $e_1$ in process A causes two events $e_3$ and $e_5$, which can each occur over a range of times such that 
\begin{flalign*}
t_3 - t_5 \geq -W_1 \quad (W_1 > 0) 
\end{flalign*}
thereby forming a timing fork. 
In other words, the minimum separation between the occurrence times of the two daughter events is lower bounded by a number:
\begin{flalign*}
\inf (t_3) - \sup (t_5) = -W_1 
\end{flalign*}
Similarly, for process C:
\begin{flalign*}
t_6 - t_4 \geq W_2 > 0, \text{where} \\
\inf (t_6) - \sup (t_4) = W_2 
\end{flalign*}

Note that nothing about this formulation requires events $e_1$ and $e_5$ to be on the same process, only that $e_5$ is caused by $e_1$, and hence occurs at a later time. Fig. \ref{tzz_exp} is depicted this way to tie in better with later sections. The same holds for $e_2$ and $e_6$. Now, under a particular case, which is when $e_4$ occurs after $e_3$ ($t_4-t_3\geq0$) and $W_2 - W_1 > 0$, we can determine an ordering of the events $e_5$ and $e_6$ as follows:
\begin{flalign*}
(t_6 - t_5) - (t_4 - t_3) &\geq W_2 - W_1 \\
\implies (t_6 - t_5) &\geq W_2 - W_1 \\
\implies t_6 - t_5 &> 0
\end{flalign*}
This type of timing constraint, where combining multiple timing forks appropriately allows for determining ordering on events that are not connected by a single point of divergence, is called a timing zigzag. The quantity $W_2 - W_1$ is referred to as the weight of the zigzag.
The crucial point to understand is that there is no common event, no single point of divergence, that determines the time of events $e_1$ and $e_2$, and that they are completely independent of each other. In other words, this case is fundamentally different from a simple timing fork and cannot be reduced to one. In fact, we did not even assume an ordering on the events $e_1$ and $e_2$ in our analysis above. Despite this, there is timing information that is not a simple "a-before-b" relation, that can be inferred based on other timing information.  

Further, this method of combination can be easily extended to create a zigzag with any number of constituent timing forks. This can lead to significantly more detailed, higher-order information about the timing of sets of events that are seemingly unrelated. 

In the description that follows, we make, to the best of our knowledge, the first known use of zigzag causality \cite{dan2017using,manohar2023timed}, in circuit design. In particular, we apply this to solve the classic mutual exclusion problem.

\section{Opportunistic Mutual Exclusion}

In the conventional mutual exclusion setup, there is a shared resource that needs to be used by at most one client at a time, clients which compete for the use of this resource, and a server, which holds a token which determines who is allowed to use the resource. The server hands out the token to one of the clients that made a request, making a decision arbitrarily. The client returns the token once it is done using the resource. When the server receives the token, it can now make the next decision on which client the resource should be allocated to. This process repeats, possibly forever. 

\begin{figure*}[t]
\centerline{\includegraphics{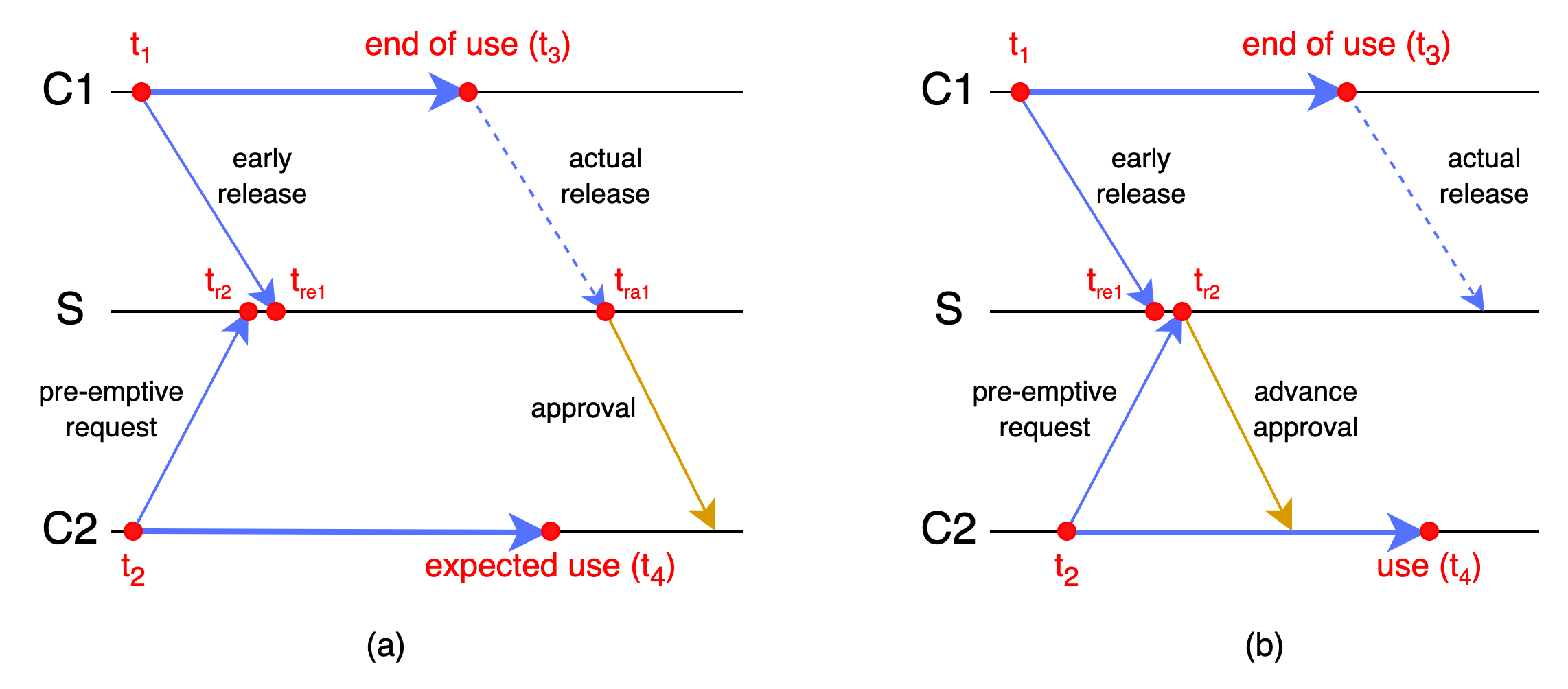}}
\caption{Timing Zigzag. (a) The standard case where the request from the second process arrives too early to exploit the timing zigzag. (b) The interesting case where the request from the second process arrives in between the early release and actual release of the first process.}
\label{tzz}
\end{figure*}

Consider the scenario shown in Fig. \ref{tzz}, where C1 and C2 are the two processes making requests to a server, S, in order to use a shared resource (not shown). In the most general case, we do not assume any a priori knowledge on when each process is going to request for use, or stop using and release the shared resource. 

In the scenario we describe here, suppose the server had knowledge of the following:
    \begin{enumerate}
        \item \textbf{Early Release Time}: The time ($t_3-t_1$) prior to the actual cessation of use of the resource, that C1 informs the server. 
        \item \textbf{Pre-emption Time}: The time ($t_4-t_2$) prior to requiring the resource, that C2 sends a request to the server. 
        \item \textbf{Link Delay}: The delays on the wires between C1, C2 and S. 
    \end{enumerate}

Once again, the definitions above are actually intervals, and when we say the server has this knowledge, we mean that it knows the bounds on these time intervals. With this information, S can calculate the upper bound, $W_1$ on $t_3-t_{re1}$ and the lower 
bound $W_2$ on $t_4-t_{r2}$. If $W_2 \geq W_1$, then there is something interesting that the server can do. We call this the asymmetric case, since the complementary bounds on the times when C2 releases early, and C1 requests preemptively are not known. If they are known, then we are in the symmetric case. 

Now, suppose C1 is using the resource and C2 places a request while the resource is still in use. If the request from C2 arrives before the early release, as shown in Fig. \ref{tzz}(a), then the server cannot ensure mutual exclusion. Since it has no internal clock, there is no measure of \textit{how} early the preemptive request arrived. So, the server must wait for the actual release from C1 to know that the resource is free, and only then grant approval to C2.

\begin{figure}
\centerline{\includegraphics[scale=0.9]{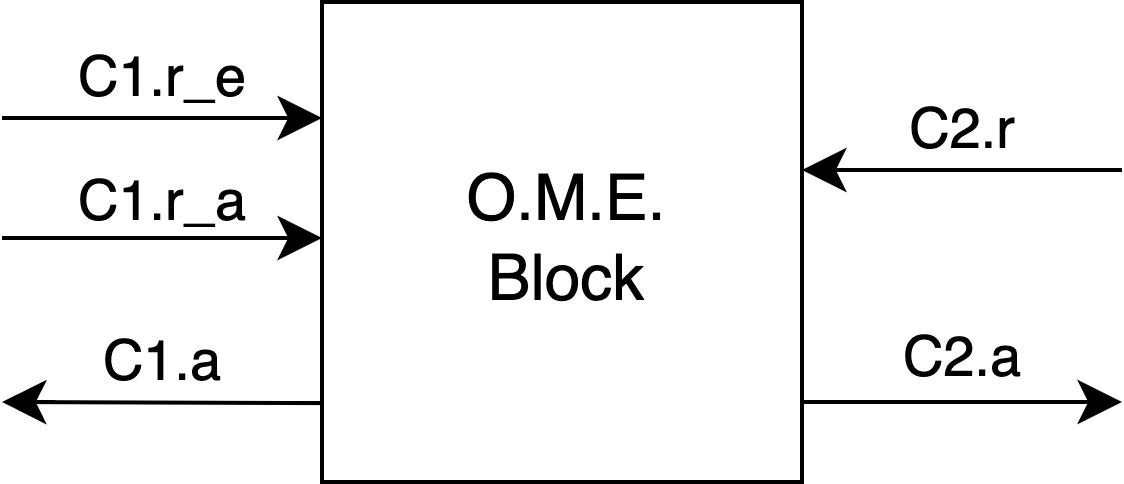}}
\caption{Opportunistic Mutual Exclusion Block - Asymmetric Case}
\label{ome_asym}
\end{figure}

The important physical difference in the channels, as shown in Fig. \ref{ome_asym}, is that C1 must have two request wires coming in to the server, the early ($r_e$) and actual ($r_a$), both of which are raised when requesting the resource. When releasing the resource, C1 lowers the early wire first, according to the early release time constraint above to signal that it is `almost done'. Then, when it is finally done with using the resource, it lowers the actual wire as well. 

The interesting case is if the request from C2 arrives after the early release from C1. In this case, since we know that the zigzag has positive weight ($W_2 - W_1 \geq 0$), the server knows that even if it grants the approval immediately, the earliest time at which the resource will be used is later than the latest time at which the resource will be released. Hence, it grants the advance approval, as shown in Fig. \ref{tzz}(b). This results in a reduction of the time for which the resource is idle, as opposed to the usual scenario when the server must wait for explicit information about the end of use of the resource to reach it. Note that, for checking if this advance approval is legal, the server does not need to know the relative times between $t_{r2}$ and $t_{re1}$, only the order in which they occurred, voiding the necessity for an internal clock. We call this \textit{opportunistic} mutual exclusion, since the necessary ordering of events that needs to occur can only be known at runtime. If the interesting case does occur, the server can `opportunistically' grant access to the other resource before revoking access from the first one, without violating the actual mutual exclusion constraint. 

\begin{figure}[t]
\centerline{\includegraphics[scale=0.9]{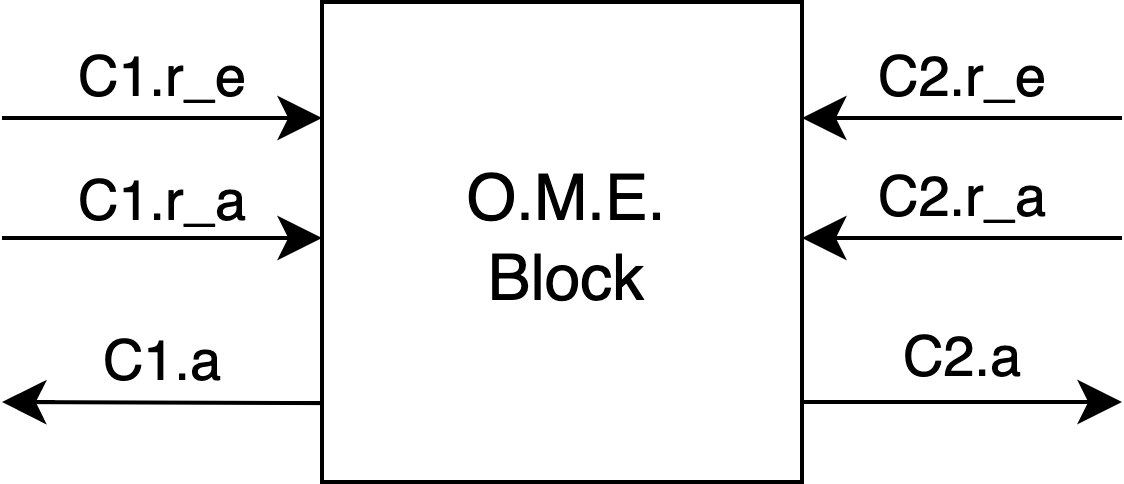}}
\caption{Opportunistic Mutual Exclusion Block - Symmetric Case}
\label{ome_sym}
\end{figure}

The symmetric case is quite similar, with both channels needing the early and actual request wires, as in Fig. \ref{ome_sym}. For this to happen, we need two distinct zigzags to have positive weight. \textit{The two zigzags are not inter-dependent.} One relates the early release time of C1 with the pre-emption time of C2, and the other relates the early release time of C2 with the pre-emption time of C1. In effect, each of C1 and C2 have two independent time variables that they can independently determine, which may result in zero or more zigzags having positive weight. In this case, the designer has additional freedom to decide what would count as a pre-emptive request. In the asymmetric case, C2 only had one request wire to raise in order to potentially receive the opportunistic grant. Here, the server could require that it raise both wires to be considered for the opportunistic case, or require only one. In the implementation described later, we assume the former. The circuit for both cases can be augmented with just a few gates so that the opportunistic mode can be turned on or off with a single bit.

Once again, the behavior described above cannot be captured by a timing fork, since the behavior of C1 and C2 are not coupled in any way, and a zigzag is needed in order to obtain any useful timing information.

\section{Three Arbiter Method}

The straightforward implementation of the asymmetric case described in the previous section, requiring only one internal state variable, is shown below. Refer to \nameref{sec:appendix} for details about the CHP notation.

\begin{hse}
x=0;
*[[\|C1.r`e&C1.r`a->C1.a+
   []   \; x|C2.r->C2.a+
  \|];
  [~C2.a->
     [\|~C1.r`e->
       [\|~C1.r`a->C1.a-
        []  C2.r->C2.a+;x+;[~C1.r`a];C1.a-
       \|]
      [] \;C2.r->[~C1.r`a];C1.a-
     \|]
  []~C1.a->[~C2.r];C2.a-;x-
  ]
 ]

\end{hse}

The first arbitration chooses between the two requesters in the idle state of the resource and grants approval in the form of an acknowledge signal, based on an arbitration. Since this is the asymmetric case, if C2 was granted access, there is nothing interesting to be done and the server just waits for the handshake to be completed and returns to the beginning. However, if C1 was granted access, then the server enters the second arbitration. Here, there are three cases: two standard and one interesting. The two standard ones are:
\begin{itemize}
    \item C2 requests before the early release.
    \item C1 finishes using before C2 even requests.
\end{itemize}
In the first case, we cannot exploit the zigzag since it is possible that mutual exclusion will be violated if approval is granted to C2. It is pertinent to remember that the server does not have any information about `how early' C2 requested before the early release and thus cannot exploit the zigzag. In the second case, C1 completes before C2 requests, so there is nothing interesting to be done, either. 

Finally, the interesting case, which the circuit is designed to exploit, occurs when:
\begin{itemize}
    \item C2 requests between the early release and actual release of C1.
\end{itemize}
In this case, the server grants approval to C2 and only then completes the handshake with C1, whenever C1 performs the actual release. The state variable $x$ is used to bypass the first arbitration whenever approval was granted to C2 in this manner, so that the handshake can be completed. Once this is done, the system is back in the quiescent state, and the process repeats.

In this implementation, three non-deterministic selections are required, resulting in three arbiters. This can be quite expensive during realization.

Next we present an alternative method that reduces the number of arbiters. The handshaking expansion described initially looks like it would require two arbiters but we show that it can be reduced to a single arbiter.  

\section{Single Arbiter Method}

In the alternate implementation below, we require two state variables, but only two non-deterministic selections, instead of three. The second process performs the initial arbitration. As before, granting approval to C2 first does not result in any interesting cases. If approval is granted to C1 and C2 requests too early, the handshake with C1 is completed before proceeding. The purpose of the first process is to complete the handshake with C1 in parallel with the other operations, whenever the state variable $g$ is set. If C2 requests between the early and actual release, then the C1 handshake is allowed to complete in parallel with approving C2. In essence, this is a parallel form of the method described earlier, with three arbiters.  

\begin{hse}
g=0;
*[[g&~C1.r`a&~C1.r`e];C1.a-;g-]
\pll
*[[\|~g&C1.r`e->[C1.r`a];C1.a+;
                    [\|  C2.r->g+;[~g]
                     []~C1.r`e->g+
                    \|]
  []      C2.r->C2.a+;[~g&~C2.r];C2.a-
 \|]
]
\end{hse}

Now, notice that we can actually rewrite the nested non-deterministic selections as a single 4-way non-deterministic selection, with the introduction of a new state variable to distinguish between the two different `modes' of the selection. This results in the following version:

\begin{hse}
g=0;f=0;
*[[g&~C1.r`a&~C1.r`e];C1.a-;g-]
\pll
*[
[\| ~f&~g&C1.r`e->[C1.r`a];C1.a+;f+
 []      ~f&C2.r->C2.a+;[~g&~C2.r];C2.a-
 []       \,f&C2.r->g+;[~g];f-
 []     \,f&~C1.r`e->g+;f-
\|]
]
\end{hse}

Finally, impelled by the facts that the 4-way arbitration actually has two two-element sets of disjoint guards and that the two sets of guards are actually quite similar, we attempt to decompose this into a single arbiter. If the `mode' of the arbiter is determined by $f$, then one guard of the arbiter is trivial, it is just $C2.r$. For the other, notice that if the fourth guard had a $\neg g$ term in conjunction, then the first and fourth guards could be combined. Observing that $g$ is always low when entering the `$f$-mode' of the arbiter, we can see that adding a $\neg g$ term to this guard is a valid strengthening. After doing this, the second guard to the single arbiter would reduce to $G = \neg g \land (f \oplus C1.r_e)$. 

Hence, we can decompose the arbitration in the process to a simple arbiter that operates in two different modes, as shown below. This is the final form that we convert to production rules to implement the circuit.  

\begin{hse}
g=0;f=0;
*[[g&~C1.r`a&~C1.r`e];C1.a-;g-]
\pll
*[[\|  \,G->v+;[~G];v-
   []C2.r->u+;[~C2.r];u-
  \|]]
\pll
*[[ ~f&v->[C1.r`a];C1.a+;f+
  [] ~f&u->C2.a+;[~g&~C2.r];C2.a-
  []  \,f&u->g+;[~g];f-
  []  \,f&v->g+;f-
 ]]
\end{hse}

It is trivial to check that the 4-way selection is now deterministic, since $u$ and $v$ are arbiter outputs and only one of them can be high at any time.

\begin{figure*}[t]
\centerline{\includegraphics[scale=0.5]{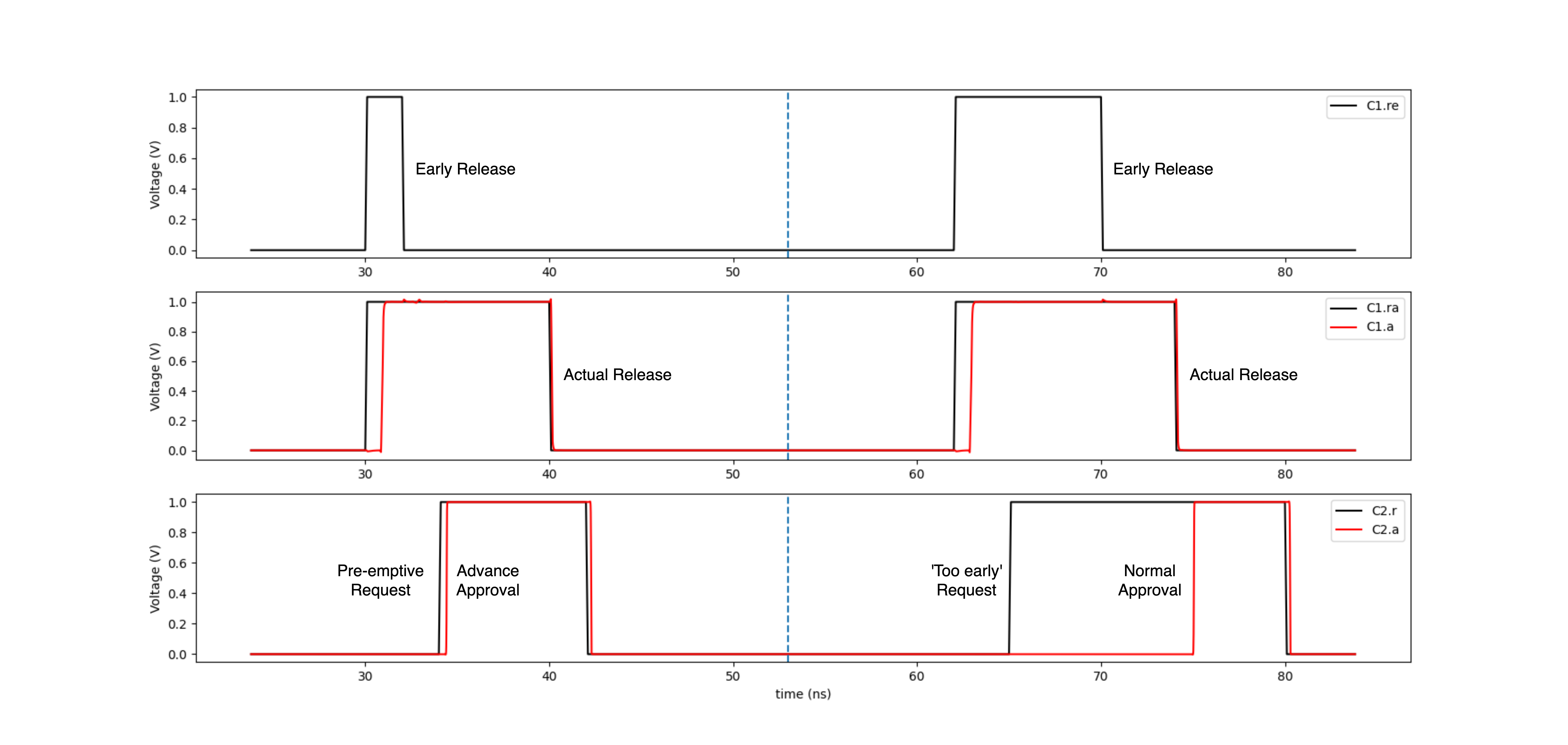}}
\caption{Waveforms from a SPICE simulation of the circuit for the asymmetric case. Requests, generated by the environment, are shown in black. Acknowledges, generated by the circuit, are shown in red. In the case to the left of the dotted line, both requests on C1 are asserted, followed by the corresponding acknowledge. Then, the early request of C1 is deasserted (top), and the C2 request (bottom) is asserted before the actual request of C1 (middle) is deasserted, resulting in the advance approval described above. In the case to the right of the dotted line, both requests on C1 are asserted, followed by the corresponding acknowledge. Then, before the early request of C1 can be deasserted, the C2 request is asserted. Now, the server must wait for C1 to fully complete before acknowledging C2. 133fJ was consumed over one sequence of handshakes (30ns - 45ns).}
\label{waves_asym}
\end{figure*}

\begin{figure*}[t]
\centerline{\includegraphics[scale=0.27]{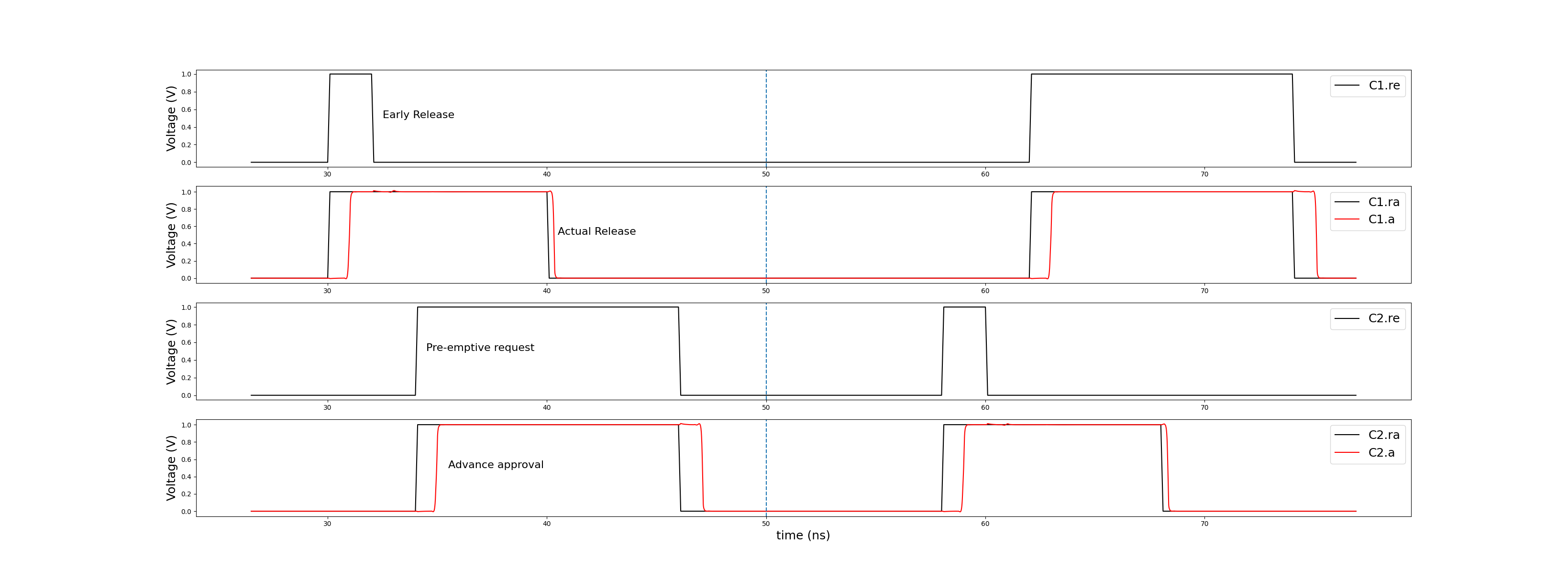}}
\caption{Waveforms from a SPICE simulation of the circuit for the symmetric case. Requests, generated by the environment, are shown in black. Acknowledges, generated by the circuit, are shown in red. To the left of the dotted line is the case described earlier in the asymmetric circuit, with C2 receiving the opportunistic grant. To the right is its symmetric counterpart, where C1 receives the opportunistic grant. 201fJ was consumed over one sequence of handshakes (30ns - 45ns).}
\label{waves_sym}
\end{figure*}

\section{Symmetric Case}
The symmetric case, described briefly earlier, occurs when both zigzags have positive weights, i.e. if there is a possibility of C2 requesting in between early and actual release of C1, and, a possibility of C1 requesting in between early and actual release of C2, and mutual exclusion can be guaranteed in both cases. The resulting circuit has the following handshake expansion, which can be derived quite easily from the asymmetric case. The 6-way non-deterministic selection that exists now is an extension of the 4-way selection before, with the additional two cases being the symmetric counterpart of the `$f$-mode'. 

\begin{hse}
g1=0;g2=0;f1=0;f2=0;
*[[g1&~C1.r`a&~C1.r`e];C1.a-;g1-]
\pll
*[[g2&~C2.r`a&~C2.r`e];C2.a-;g2-]
\pll
*[
[\| ~f1&~f2&~g1&C1.r`e->[C1.r`a];C1.a+;f1+
 [] ~f1&~f2&~g1&C2.r`e->[C2.r`a];C2.a+;f2+
 []   f1&~f2&~g2&C2.r->g1+;[~g1];f1-
 []      \;f1&~f2&~C1.r`e->g1+;f1-
 []   ~f1&f2&~g1&C1.r->g2+;[~g2];f2-
 []      \;~f1&f2&~C2.r`e->g2+;f2-
\|]
]
\end{hse}

As before, this can be decomposed into a single arbiter operating in 3 modes, determined by $f1$ and $f2$. We use the shorthand @Cx.r = Cx.r`a & Cx.r`e@ in the third and fifth guards.

\section{Circuit Design and Evaluation}
\label{sec:evaluation}

We realized the final circuits described above, in the TSMC 65nm technology node, and performed simulations using Xyce \cite{keiter2022xyce}. All simulations were done at the schematic level. We include energy metrics for the sake of completeness. The resulting design for the asymmetric case consumed an energy of 133fJ for a sequence of two handshakes, one on each channel. The total leakage power was 876nW. We measured the response delays on the C1 and C2 ports, with pulse sources generating requests. Under this test condition, the asymmetric circuit exhibited a 330ps delay on the C2 port, and a 910ps delay on the C1 port. The difference is due to the asymmetry in the circuit and channel models. The sequence of transitions for C1 is longer than C2, due to the additional computation to setup the circuit to respond to a possible early request from C2. 

The corresponding energy per sequence of handshakes and leakage power of the circuit for the symmetric case were 201fJ, and 1.61$\mu$W respectively. The delays on both ports were 1.05ns, which, as expected, are slightly longer due to the more complex circuit.

The SPICE simulation results are shown in Figs. \ref{waves_asym} and \ref{waves_sym}. Only the relevant input and output variables are shown, with the inputs to the circuits (requests) in black and the outputs generated by the circuit (acknowledges) in red. The periods where the two acknowledges overlap are the latency reductions that are obtained by exploiting the timing zigzag. 

Though there is an added delay of about 600ps in the asymmetric case, the expectation when using this circuit is that the interesting case, where the zigzag can be exploited, occurs frequently. The weight of the zigzag, $W_2-W_1$, is actually unbounded, since it is determined by the C1 and C2 processes providing the server information about their usage time of the shared resource. This could be orders of magnitude larger than the delay of this circuit, which can result in significant latency benefits. Comparing the symmetric circuit to the baseline 330ps delay, we see that  $\sim$700ps of additional delay is incurred. 

These delays can be used to determine if the overhead of the opportunistic mutual exclusion mechanism is offset by the gains obtained by exploiting the timing zigzag using either the symmetric or asymmetric case.


\section{Discussion}

The circuit described in this article is significantly more complex than a simple arbiter that handles mutual exclusion between two clients. However, in cases where certain timing information is known, the additional hardware cost is worth the reduction in idle time of the critical resource. 

The final reduction to the form that uses only one arbiter has one disadvantage that is not present in the cases with three/two arbiters. Decomposing the 4-way non-deterministic selection into a single arbiter is not without a cost. When using an arbiter with guards that are modified between one use and the next, there is a possibility of instability during the switching. To see this, consider the following: the switching of modes (change of guards) in the arbiter must be caused by a transition on some control variable which, in this case, is $f$. Since it appears in more than one process, $f$ is a now a shared variable. In the conventional use of shared variables, there is communication between two processes to ensure that the two do not attempt to modify/read this variable at the same time. This ensures that the variable has the correct value when a process accesses it. However, in our circuit, once $f$ is asserted, there is no possible transition in the arbiter that can be sent back to the main process to acknowledge the fact that the change of guard has been completed. The arbiter only has three possible output variables, $u$, $v$ and $G$. But none of them have uniquely defined values at the end of the $f$ transition, since they all depend on the values of the external variables directly. Hence, the main process \textit{cannot} know whether the change of guard has been successfully completed, without knowing the delay of the gates in the arbiter.

As explained above, this instability is actually fundamental when `overloading' an arbiter to perform multiple arbitrations in sequence. Though this is a problem in the simulation scenario where all gates delays are randomized, in practice, this can be resolved by an elementary timing assumption: requiring that the gate that implements the boolean function for $G$ be faster than the sequence of transitions from @f+@ to either of the following @f-@ transitions---both of which include a change of the handshake variables in the opportunistic mutual exclusion circuit followed by a response from the environment. Through the use of additional state variables, this assumption can be reduced to requiring that a single inverter be faster than a few more complex gates, which is easily achievable in practice. Our evaluation in Section~\ref{sec:evaluation} corresponds to a circuit that includes these additional state variables as part of the overhead reported.

Finally, since we are already making use of timing constraints in order to even decide whether to use this circuit, this additional local constraint, which can be guaranteed by design, does not pose significant restriction.

\section{Conclusion}
In this article, we presented a novel method to exploit certain subtle timing constraints in order to design mutual exclusion servers. We showed that these servers can reduce the idle time of a shared resource by opportunistically granting access to more than one client at a time, based on knowledge of the expected time when the clients will stop and start using the resource. We described handshaking expansions for the processes, implemented the same in the TSMC 65nm node and evaluated the performance. Finally, we discussed places where using this circuit is actually warranted, its advantages and possible drawbacks. All of this demonstrates the value of using the notion of zigzag causality, originally introduced in \cite{dan2017using}, in circuit design. To our knowledge, this is the first work to do so. 

\section*{Acknowledgments}
This work was supported in part by DARPA IDEA grant FA8650-18-2-7850, and in part by DARPA POSH grant HR001117S0054-FP-042. Yoram Moses is the Israel Pollak academic chair at the Technion, and was supported in part by the BSF grant 2015820 which is coincident with NSF-BSF grant CCF 1617945.

\bibliography{refs}
\bibliographystyle{ieeetr}

\section*{Appendix I}
\label{sec:appendix}

Communicating Hardware Processes (CHP) is a hardware description language used to describe clockless circuits derived from C.A.R. Hoare’s Communicating Sequential Processes (CSP) \cite{hoare1978csp}. A full description of CHP and its semantics can be found in \cite{martin1991synthesis}. Below is an informal description of a subset of that notation that we use, listed in descending precedence, replicated from \cite{maybeex}. For a complete discussion of the interaction between the handshake expansions of channel actions like send and receive and the composition operators, see \cite{manohar2001analysis}.

A \textbf{Channel} X consists of a \textbf{request} X.r and either an \textbf{acknowledge} X.a or \textbf{enable} X.e. The acknowledge and enable serve the same purpose, but have inverted sense. With these signals, a channel implements a network protocol to transmit data from one QDI process to another.
\begin{itemize}
    \item \textbf{Skip}: @skip@ does nothing and continues to the next command.
    \item \textbf{Dataless Assignment}: @n+@ sets the node @n@ to @true@ and @n-@ sets it to @false@.
    \item \textbf{Probe}: @X?@ is used determine if the channel is ready for a receive action, returning the value waiting on the request @X.r@ without executing the receive. @X!@ is used to determine if the channel is ready for a send action, expanding into either @~X.a@ given an acknowledge or @X.e@ given an enable. For dataless channels, the syntax is simplified to @X@.
    \item \textbf{Sequential Composition}: @S;T@ executes the programs @S@ followed by @T@.
    \item \textbf{Parallel Composition}: @S\|\|T@ executes the programs @S@ and @T@ in any order.
    \item \textbf{Deterministic Selection}: @[G`1->S`1[]...[]G`n->S`n]@ where @G`{i}@, called a guard, is a dataless expression and @S`{i}@ is a program. The selection waits until one of the guards, @G`{i}@, evaluates to @true@, then executes the corresponding program, @S`{i}@. The guards must be stable and mutually exclusive. The notation @[G]@ is shorthand for @[G->skip]@, which corresponds to waiting for @G@ to become true. 
    \item \textbf{Non-Deterministic Selection}: @[\|G`1->S`1[]...[]G`n->S`n\|]@ is the same as Deterministic Selection except that the guards do not have to be stable or mutually exclusive. If two or more evaluate to @true@ simultaneously, then one is picked arbitrarily (not necessarily random). In a circuit, this choice is implemented by a collection of arbiters and synchronizers. When two or more guards evaluate to @true@ simultaneously, it can cause a metastable state in the arbiter or synchronizer. This metastable state then resolves non-deterministically, giving the grant to one of the branches of the selection statement. Therefore, the digital model of this selection statement is also non-deterministic in such a condition.
    \item \textbf{Repetition}: @*[G`1->S`1[]...[]G`n->S`n]@ is similar to the selection statements. However, the action is repeated until no guard evaluates to @true@. @*[S]@ is shorthand for @[true->S]@.
\end{itemize}

\section*{Appendix II}
The CMOS-implementable production rules set for the server process (asymmetric case) is shown below. A pair of production rules jointly define the combinational/state-holding gate that implements that variable. Variables with an underscore-prefix are typically used to represent inverted forms of variables. Combinational gates are labeled with (*) and hence the opposing production rule for that variable is omitted. The production rules for $u,v$ were implemented using an arbiter.

\begin{prs}
~Reset -> _Reset+ \;(*)
~_Reset | ~C1.r`e -> _C1.r`e+ \;(*)

~u`{reg1} -> _u`{reg1}+ \;(*)
~u`{reg2} -> _u`{reg2}+ \;(*)
~v`{reg1} -> _v`{reg1}+ \;(*)
~v`{reg2} -> _v`{reg2}+ \;(*)

_u`{reg1} & _u`{reg2} & _v`{reg1} & _v`{reg2} -> reg- \;(*)

@~reg & ~g \;&@
((~_f & ~C1.r`e) | (~f & ~_C1.r`e))  -> G`{arb}+ 
Reset | reg & (v`{reg1} | v`{reg2})  -> G`{arb}- 

\end{prs}
$ \qquad u,v = Arbiter (G_{arb}, C2.r)$
\begin{prs}

~u -> _u+ \;(*) 
~v -> _v+ \;(*)
    
(~f & ~C1.a & ~g) -> _g`{reg}+ 
C1.a & v`{reg1} & reg -> _g`{reg}-
~_g`{reg} -> \;\;g`{reg}+ \;(*)

~_u & ~_f & ~reg                   -> u`{reg1}+ 
Reset | (reg & _g & _f & _C1.a)    -> u`{reg1}- 

\end{prs}
\begin{prs}

~_v & ~_f & ~reg                   -> v`{reg1}+ 
Reset | (reg & _f & _v)            -> v`{reg1}- 

~_u & ~f & ~reg                   -> u`{reg2}+ 
Reset | (reg & _g & _u)           -> u`{reg2}- 

~_v & ~f & ~reg & ~g`{reg}         -> v`{reg2}+ 
Reset | (reg & C1.a & f & _v)      -> v`{reg2}- 

~_Reset | (~C1.a & ~f)             -> _g+ 
((reg & u`{reg1}) | g`{reg}) & f   -> _g- 
~_g -> \;\;g+ \;(*)

@~_Reset \;|@ 
((~_u`{reg1} & ~C1.a) | ~_v`{reg1}) & ~_g -> _f+
reg & v`{reg2} & C1.a -> _f- 
~_f -> \;\;f+
Reset | (g & _f) -> \;\;f-

@~_Reset | (~v`{reg2} \;& @
(~_u`{reg1} | ~_g`{reg}) & ~C1.r`a & ~C1.r`e) -> _C1.a+ 
_g`{reg} & C1.r`a & C1.r`e & v`{reg2} & reg -> _C1.a- 
~_C1.r`e & ~_C1.a -> \;\;C1.a+ 
Reset | (_C1.r`e & _C1.a)  -> \;\;C1.a- 

~C2.r  & ~u`{reg2}  -> _C2.a+
reg & u`{reg2} & C2.r -> _C2.a- 
~_C2.a -> \;\;C2.a+ \;(*)
\end{prs}



\end{document}